\documentclass[a4paper,11pt]{article}

\def \t 	{\boldsymbol{\hat t}}
\def \n 	{\boldsymbol{\hat n}}
\def \r		{\boldsymbol{r}}
\def \u		{\boldsymbol{u}}

\def \G		{\boldsymbol{G}}
\def \f		{\boldsymbol{f}}
\def \v		{\boldsymbol{v}}

\def \Gammavec		{\boldsymbol{\Gamma}}
\def \F		{\boldsymbol{F}}
\def \e		{\rm e}

\def \zero	{\boldsymbol{0}}
\def \zper 	{\zeta_{\bot}}
\def \zpar 	{\zeta_{\parallel}}
\def \d 	{\mathrm{d}}

\usepackage{graphicx}  
\usepackage{bm}        
\usepackage{amssymb,amsfonts,amsmath}   
\usepackage[margin=1.2in]{geometry}

\begin{document}

\title{Spontaneous Oscillations of Elastic Filaments Induced by Molecular Motors}

\author{Gabriele De Canio, Eric Lauga, Raymond E. Goldstein \\ 
{\small \it Department of Applied Mathematics and Theoretical Physics,} \\
{\small \it Centre for Mathematical Sciences, University of Cambridge,} \\ 
{\small \it Wilberforce Road, Cambridge CB3 0WA, United Kingdom}}

\date{\today}

\maketitle

\begin{abstract}
It is known from the wave-like motion of microtubules in motility assays that the piconewton 
forces that motors produce can be sufficient to bend the filaments. In cellular phenomena 
such as cytosplasmic streaming, molecular motors translocate along cytoskeletal filaments, 
carrying cargo which entrains fluid.   When large numbers of such forced filaments interact 
through the surrounding fluid, as in particular stages of oocyte development in {\it Drosophila
melanogaster}, complex dynamics are observed, but the detailed mechanics underlying 
them has remained unclear.  Motivated by these observations, we study here perhaps 
the simplest model for these phenomena: an elastic filament, pinned at one end, acted on by 
a molecular motor treated as a point force.  Because the force acts tangential to the filament, 
no matter what its shape, this `follower-force' problem is intrinsically non-variational, and thereby 
differs fundamentally from Euler buckling, where the force has a fixed direction, and which, 
in the low Reynolds number regime, ultimately leads to a stationary, energy-minimizing shape.  
Through a combination of linear stability theory,  analytical study of a solvable simplified 
`two-link' model, and numerical studies of the full elastohydrodynamic equations of motion 
we elucidate the Hopf bifurcation that occurs with increasing forcing of a filament, 
leading to flapping motion analogous to the high Reynolds number oscillations of a garden hose 
with a free end.
\end{abstract}

\section{Introduction}

Motor protein translocation along cytoskeletal filaments within eukaryotic cells, a phenomenon 
which is central to many aspects of physiology and development, underlies one of the most 
fundamental examples of ``fluid-structure'' interactions in cellular biology: the phenomenon of
cytoplasmic streaming.  Discovered first in aquatic plants in 1774 by Bonaventura Corti \cite{Corti}, 
it is now known to take place in a broad spectrum of aquatic and terrestrial organisms 
\cite{streaming_jrsi}.  In each case of motor protein-filament pairs --typically myosin-actin in 
plants and kinesin-microtubules in animals-- cargo carried along by the motors entrains 
cytoplasmic fluid, creating flows whose degree of organization reflects the architecture of 
the filament network. While in mature plants the filaments tend to be anchored along the interior 
cell wall, in young developing plant cells, and also in mature cells whose cytoskeleton has been 
transiently chemically disrupted, there is strong evidence for a self-organization processes 
\cite{Wasteneys} which likely involves filament buckling and alignment by the very flows 
created by the moving motors \cite{WoodhousePNAS}.  In the case of animals, the paradigm 
is oogenesis in the fruit fly \textit{Drosophila} \cite{oogenesis_ref}, in which a dense network of
microtubules emanates from the entire periphery of the oocyte, so that one end of each filament is 
anchored at the oocyte boundary while the distant end is free within the cellular interior. Direct
visualizations \cite{Ganguly2012} of the streaming flows (by means of endogenous tracer particles) 
and the microtubules (fluorescently labelled) show that the flows are disordered on the scale of 
the oocyte and are time-dependent on the scales ranging from seconds to many minutes.  
While the long-time variation reflects changes in the composition of the cytoskeletal fluid, the 
short-term variations arise from motion of the filaments in response to the streaming flows.

In addition to these rather complex examples of filament dynamics, we recall that in the context of
``motility assays'' it has been observed that single filaments forced by carpets of motors on a surface
can undergo a variety of buckling instabilities, particularly when one end is pinned by a `defect' in the
monolayer of motors \cite{Bourdieu, Gittes}.  Although these examples of filament deformation 
induced by molecular motors are well-known and the subject of considerable recent study \cite{Young2010,IseleHolder,Chaudhuri,Gosselin}, in the situation appropriate to streaming a precise
formulation and analysis of these problems has been lacking.  Our goal here is to present 
such an analysis, focusing on the simplest example possible: a single filament hosting a 
 molecular motor, with one filament end attached to a  wall and the other free.  
The more complex multifilament problem like that observed in \textit{Drosophila} 
oocyte streaming will be discussed elsewhere \cite{decanio2017inprep}.

Unlike in motility assays \cite{Gittes}, a filament responding to the forces produced by motors 
moving along it corresponds to a motor-induced force that is always tangential to the filament.  
Known in the mechanics literature as a ``follower force'' \cite{Herrmann1964}, this type of 
problem is intrinsically different from conventional Euler buckling where opposing thrusting 
forces are applied along a fixed axis, independent of the filament configuration.  As a consequence, 
the follower-force problem is intrinsically non-variational.  Prior studies of this dynamics were 
primarily in the context of macroscopic systems, for which damping is minimal \cite{Langthjem, 
Elishakoff}.  In such systems there is a well-known \textit{flutter instability} that can occur for 
sufficient forcing.  This idea has recently been incorporated into a model for eukaryotic flagellar 
motion \cite{BaylyDutcher} as a novel explanation for the origin of the beating waveform, and 
the present work is very much in the same spirit. 

In Sec. \ref{elasto_sec} we formulate the simplest low Reynolds number follower-force problem, 
in which the motor exerts a force on the filament but does not itself produce flow, and
demonstrate numerically the existence of a Hopf bifurcation when the force exceeds a finite threshold. 
This threshold is determined through a linear stability analysis in Sec. \ref{linstab_sec}.  A simplified 
`two-link' model of the kind used in inertial problems is solved in Sec. \ref{two-link_sec} to elucidate
the nature of the instability.  A generalization of the problem to include the fluid flow created by 
the molecular motor is presented in Sec. \ref{flow_sec}, and Sec. \ref{discussion_sec} is a 
discussion of  future possible extensions of the model.

\section{Elastohydrodynamics}
\label{elasto_sec}
\begin{figure}[t]
  \centering
  \includegraphics[width=0.6\textwidth]{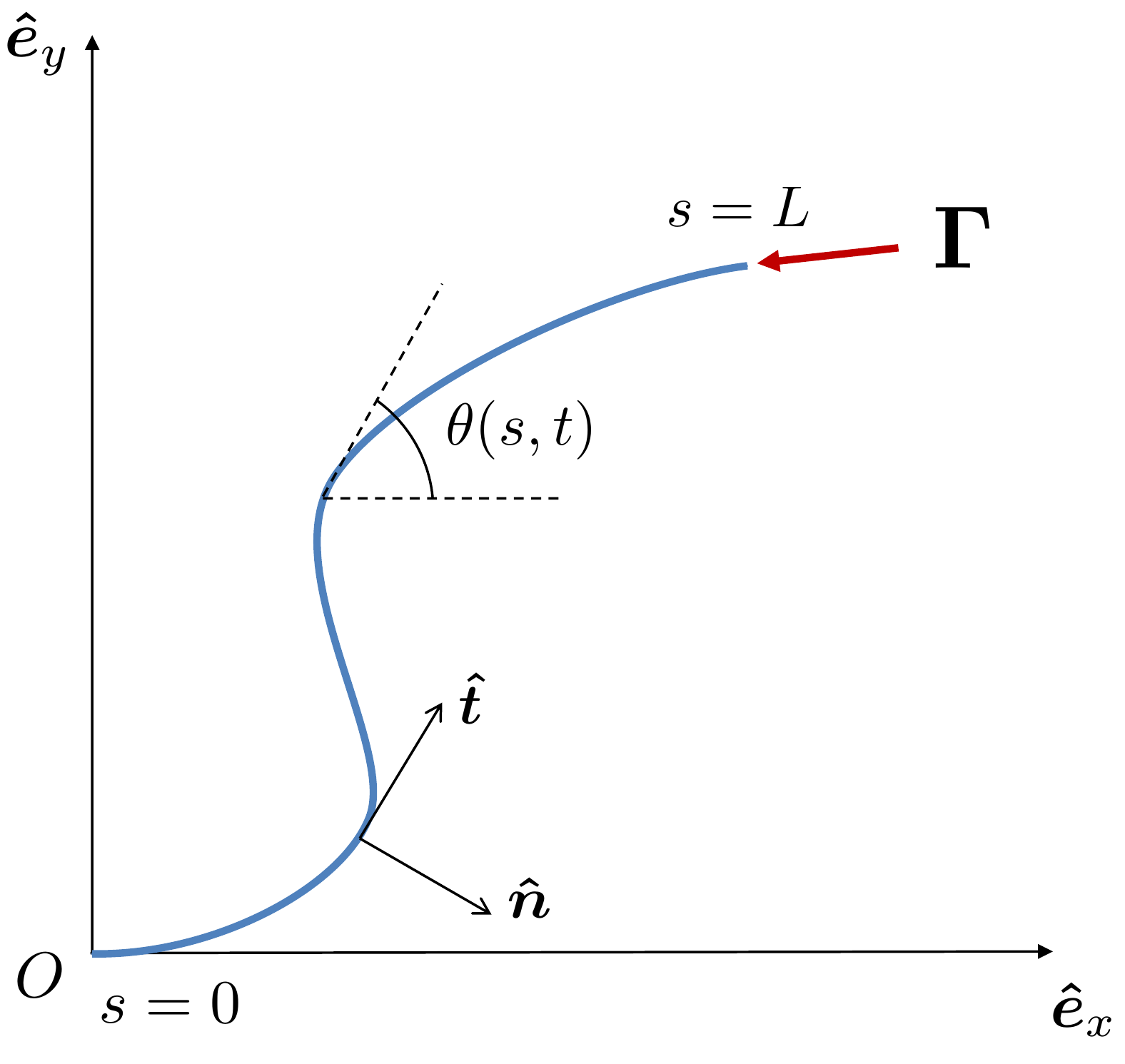}
  \caption{Schematic of a horizontal flexible filament clamped at one end with a follower force $\Gammavec$ applied 
at its tip. The filament position is defined by $\r(s,t)$, with $0\leq s\leq L$ being the arclength, or, equivalently, by 
the tangent angle $\theta(s,t)$, providing the coordinates of the clamped end.  The local tangent and unit vectors 
are $\t(s,t)$ and $\n(s,t)$, respectively.}
  \label{geom}
\end{figure}

Here we derive the low Reynolds number equations of motion for a slender elastic filament, clamped at one end and
subject to a compressive follower force $\Gammavec$, with constant magnitude $\Gamma$, moving in 
a Newtonian fluid of viscosity $\mu$, and 
confined to the plane $z=0$.  It has length $L$, diameter $b$, with $L/b\gg 1$, constant 
circular cross-section, and bending modulus $A$.  We parametrize the filament shape $\r(s,t)$ 
by its arclength $0 \leq s \leq L$ (Fig.~\ref{geom}).

\subsection{Governing equations}
We assume the standard elastic energy associated with a bent filament, expressed in terms of
its curvature $\kappa (s,t)$ as $\mathcal{E}_{\rm{el}} = \frac{A}{2} \int_0^L \kappa^2(s,t) \, \d s$, 
with a vanishing intrinsic curvature \cite{audoly2010}.  
Inextensibility is imposed through the Lagrangian multiplier $\Lambda (s,t)$ and the energy functional
associated with the local arclength conservation reads $\mathcal{E}_{\rm{ten}} = - \frac{1}{2} \int_0^L
\Lambda(s,t) \, \d s$ \cite{goldstein1995nonlinear}. 
After computing functional derivatives of the total energy, we obtain the classical elastic force per unit length 
for an inextensible filament, $\f_e$, as
\begin{equation}
\f_e=-A \r_{ssss} - (\Lambda \r_s)_s,
\end{equation}
 where subscripts indicate differentiation.  
At the clamped end we have the boundary conditions
\begin{equation}
\r(0,t) = \zero \ \ \ \ \textit{\rm and}\ \ \ \ \r_s(0,t) = \boldsymbol{\hat e}_{x} \, ,
\end{equation}
as the filament is fixed and horizontal at the clamp, while at the free end 
\begin{align}
\r_{ss}(L,t) & = \zero \, , \label{bc_tip_1}\\
 -A \r_{sss} (L,t) - \Lambda(L,t) \, & \r_s(L,t) = - \Gamma \,  \r_s(L,t) \, , \label{bc_tip_2}
\end{align}
which capture the fact the filament is torque-free and that the shearing and external force
applied at the tip must balance.
Since the follower force acts tangentially, it is {nonconservative}.  
It is this feature that gives rise to the complex dynamics in this problem.

In the Stokesian regime, the drag force acting
on the filament from the surrounding flow is classically given in the slender limit  by resistive-force theory (RFT)
\cite{gray1955propulsion,cox1970motion} which provides a local relation between the local filament velocity, $\r_t$,   
and the hydrodynamic force per unit length exerted by the surrounding fluid, $\f_h$.  When no background flow 
is present, we have 
\begin{equation}
\f_h =- \left( \zpar \t \t + \zper \n \n \right) \cdot \r_t,
\end{equation}
where $\t$ and $\n$ are the local tangent and normal unit vectors, and $\zper$, $\zpar$ (with 
 $\zper = 4 \pi \mu/[\ln(L/b)+1/2]$ \cite{pak2011} and $\zper/\zpar
\rightarrow 2$ as $L/b\to\infty$) are the drag coefficients in the perpendicular and parallel direction,
respectively
\cite{gray1955propulsion,cox1970motion}. For simplicity, we assume $\eta\equiv \zper/\zpar = 2$, 
even if more accurate 
expression can be used \cite{lighthill1976flagellar}, but for the sake of generality we write explicitly
$\eta$ throughout the paper.  While slender-body theory 
\cite{hancock1953self, cox1970motion, keller1976slender}, which consists of a more accurate treatment 
of the drag force to include nonlocal effects, could
be used, RFT has been shown 
to be a valid alternative for single filaments that are not too highly deformed, and its use significantly 
reduces the complexity of the
mathematical formulation \cite{johnson1979flagellar, goldstein1995nonlinear, wiggins1998flexive,
wiggins1998trapping, kantsler2012fluctuations, young2007stretch, becker2001instability}.

The instantaneous balance of forces for the filament is given as $\f_e+\f_h=\bf 0$, hence
\begin{equation}
-\left( \zpar \t \t + \zper \n \n \right) \cdot \r_t  - A \r_{ssss} - \left( \Lambda \r_s \right)_s =\bf 0 .
\label{forcebal}
\end{equation}
Exploiting the two-dimensional Frenet-Serret equations, $\t_s = - \kappa \, \n$ and 
$\n_s = \kappa \, \t$, this can be rewritten as
\begin{equation}
\r_t = 	\frac{1}{\zper}\left[ A \left (\kappa_{ss} - \kappa^3 \right ) + \kappa \Lambda \right] \n +
		\frac{1}{\zpar}\left( 3 A \kappa \kappa_s - \Lambda_s \right)\t \, .
\label{filconf}
\end{equation}
[Note that the form of the elastic component of the normal force often seen in the literature 
\cite{goldstein1995nonlinear},
$A(\kappa_{ss}+(1/2)\kappa^3)$, is equivalent to that in \eqref{filconf} under the redefinition of the
Lagrange multiplier: $\Lambda\to\Lambda+(3/2)A\kappa^2$.]

If we rescale lengths by $L$, time by the relaxation time $\zper L^4/A$ , and the Lagrangian multiplier 
by the elastic force $A/L^2$, then in dimensionless units Eq.~(\ref{filconf}) becomes 
\begin{equation}
\r_t = 	\left (\kappa_{ss} - \kappa^3 + \kappa \Lambda \right )  \n +
		\eta\left( 3 \kappa \kappa_s - \Lambda_s \right) \, \t \, .
\label{filconf_nondim}
\end{equation}
If we now differentiate \eqref{filconf_nondim} with respect to arclength, separate the normal and
tangent components, and notice that $\r_s \cdot \r_{ts} = 0$ to ensure local inextensibility 
($\r_s \cdot \r_s = 1$), we obtain the coupled equations describing the evolution
of the tangent angle $\theta$ and the tension, $\Lambda$, as
\begin{eqnarray}
\theta_t  &=& 	-\theta_{ssss} 
			 		 -\left[ \Lambda - 3 \left(\eta +1 \right)  \theta_s^2 \right]\, 
			 		 \theta_{ss}  - \left (\eta +1 \right) \Lambda_s \, \theta_s  \, , 
\label{theta}\\
\Lambda_{ss} - \eta^{-1}\theta_s^2 \Lambda & = &
			-\eta^{-1} \theta_s^4 
			+ 3 \theta_{ss}^2 
			+ \left( 3+\eta^{-1} \right) \theta_s \, \theta_{sss} \, , 
\label{lambda}
\end{eqnarray}
in which we have used the relation $\theta_s = \kappa$.

It is important to note that in differentiating Eq.~(\ref{filconf_nondim}) with respect to 
arclength, the boundary condition $\r(0,t) = \zero$ is lost.  To restore the missing boundary condition 
physical insight is required.  At $s=0$, the filament is not only fixed, but, trivially, it has  zero velocity, i.e.
$\r_t(0,t) = \zero$.  This consideration than leads to boundary conditions for $\theta (0,t)$ and
$\Lambda (0,t)$ when directly evaluating Eq.~(\ref{filconf}) at $s=0$.  These are 
\begin{equation}
\theta_{sss}(0,t) - \theta_s(0,t)^3 + \theta_s(0,t) \Lambda(0,t) = 0 \, ,
\end{equation} and
\begin{equation}
\Lambda_s (0,t)- 3 \, \theta_s (0,t) \, \theta_{ss} (0,t)  =  0 \, ,
\end{equation}
respectively.
The condition $\r_s(0,t)=\boldsymbol{\hat e}_{x}$ becomes $\theta(0,t) = 0 \, ,$
while Eq.~(\ref{bc_tip_1}), $\theta_{s}(1,t) = 0 \, ,$ and Eq.~(\ref{bc_tip_2}), $\theta_{ss}(1,t) = 0 \,$
and $\Lambda(1,t) = \sigma \, , \label{tension_tip}$ where
\begin{equation}
\sigma \equiv \frac{\Gamma L^2}{A} \, ,
\end{equation} 
is the dimensionless ratio between the strength of the force at the tip and the elastic force and  
is the one relevant parameter governing the dynamics of the filament.  Note that since the force 
is compressive ($\Gamma >0 $), $\sigma$ is always positive.  

\subsection{Dynamical features of a follower force}
\label{insights}
The nonvariational form of the follower force differs intrinsically from conventional Euler buckling in which the 
compressive force is always in a given direction.  Examination of the equations of motion linearized around
the straight filament, studied in much more detail in Sec. \ref{linstab_sec}, reveals important physical insights 
into the expected dynamics.  
If $y$ denotes the $y$ component of the position of the filament, the  linearized nondimensional form of 
Eq.~(\ref{filconf_nondim}) is classically given by 
\begin{equation}
y_t = - y_{xxxx} - \Lambda y_{xx} \, .
\end{equation}
If we calculate the rate of change of the (nondimensional) bulk energy
\begin{equation}
{\cal E}=\frac{1}{2}\int_0^1\!\! \left( y_{xx}^2-\Lambda y_x^2\right) \d x,
\label{E_define}
\end{equation}
then repeated integrations by parts and imposition of the boundary condition \eqref{bc_tip_2} yields the result
\begin{equation}
{\cal E}_t = -\int_0^1 \!\!  \left(y_{xxxx}+\Lambda y_{xx}\right)^2 \d x - 
\sigma  y_{t}(1) \, y_{x}(1) \, .
\label{E_evolve}
\end{equation}
The integral term is clearly negative semi-definite, and absent the final
term (as in Euler buckling) it would drive the energy monotonically downward. The boundary term arises 
from the fact that the 
follower force always acts tangentially, and it is clear that depending 
on its sign, the follower force either removes or injects energy into the system, 
possibly giving rise to persistent motion as discussed below.

\subsection{Buckling and Flapping}
\label{buckling_sec}

\begin{figure}[t]
\centering
   \includegraphics[width=.65\textwidth]{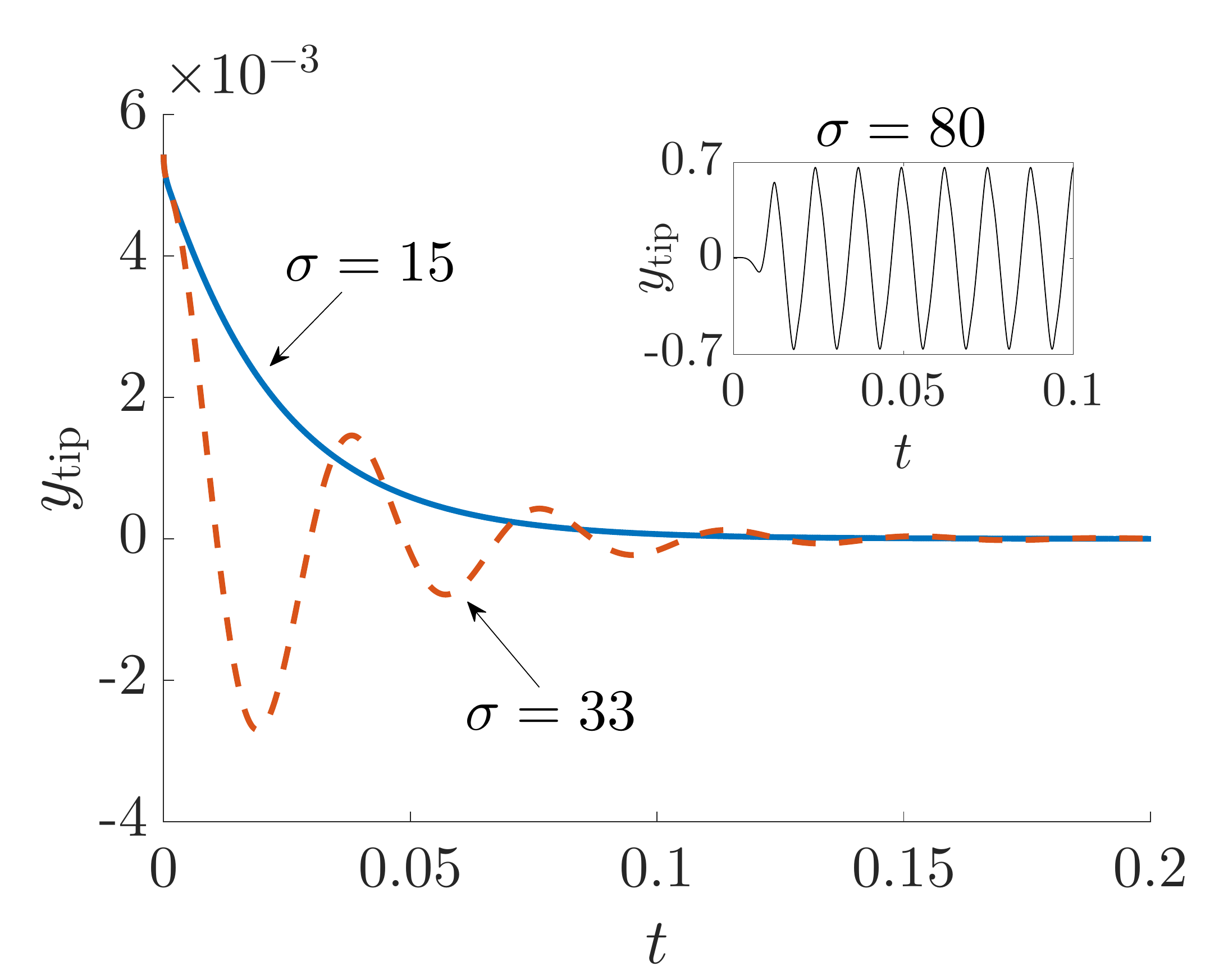}
\caption[]{Tip displacement as a function of time for three different values of $\sigma$. Blue solid line,  
$\sigma = 15$: the filament returns monotonically to its original shape. Red dashed line, $\sigma = 33$: after a
transient, the oscillation dies out as the filament straightens. Inset: for $\sigma = 80$ the filament shows 
a sustained periodic oscillation. }
\label{tip_3cases}
\end{figure}

The governing equations, Eqs. (\ref{theta}) and (\ref{lambda}) together with the corresponding boundary conditions, were
discretised using second-order centred finite differences in the bulk and one-sided differences at the edges.  
The resulting nonlinear system of algebraic equations was solved using Newton's method.  To overcome the constraint 
of the time step arising from the stiff nature of Eq.~(\ref{theta}), a backward Euler method, which is an 
implicit A-stable numerical scheme, was used \cite{iserles1996}. The equations were decoupled using the
values at the previous time step \cite{quennouz2015transport}.

Numerical results for a horizontal filament to which a small perturbation was initially introduced 
identify three different dynamical behaviors depending on the value of $\sigma$, as illustrated in Fig.~\ref{tip_3cases}.  For $\sigma \lesssim 20.4$ the 
filament returns monotonically to its original straight configuration (illustrated for $\sigma=15$ as the red dashed line).  In the interval $20.4\lesssim\sigma\lesssim 37.5$ the filament displays 
decaying oscillations (the case with $\sigma = 33$ is shown in blue solid line). Finally, above the threshold $\sigma\gtrsim 37.5$, we find that any perturbation grows and the motion settles into a 
finite-amplitude periodic oscillation (see inset of Fig.~\ref{tip_3cases} in the case $\sigma=80$).

\begin{figure*}[t]
\centering
   \includegraphics[width=\textwidth]{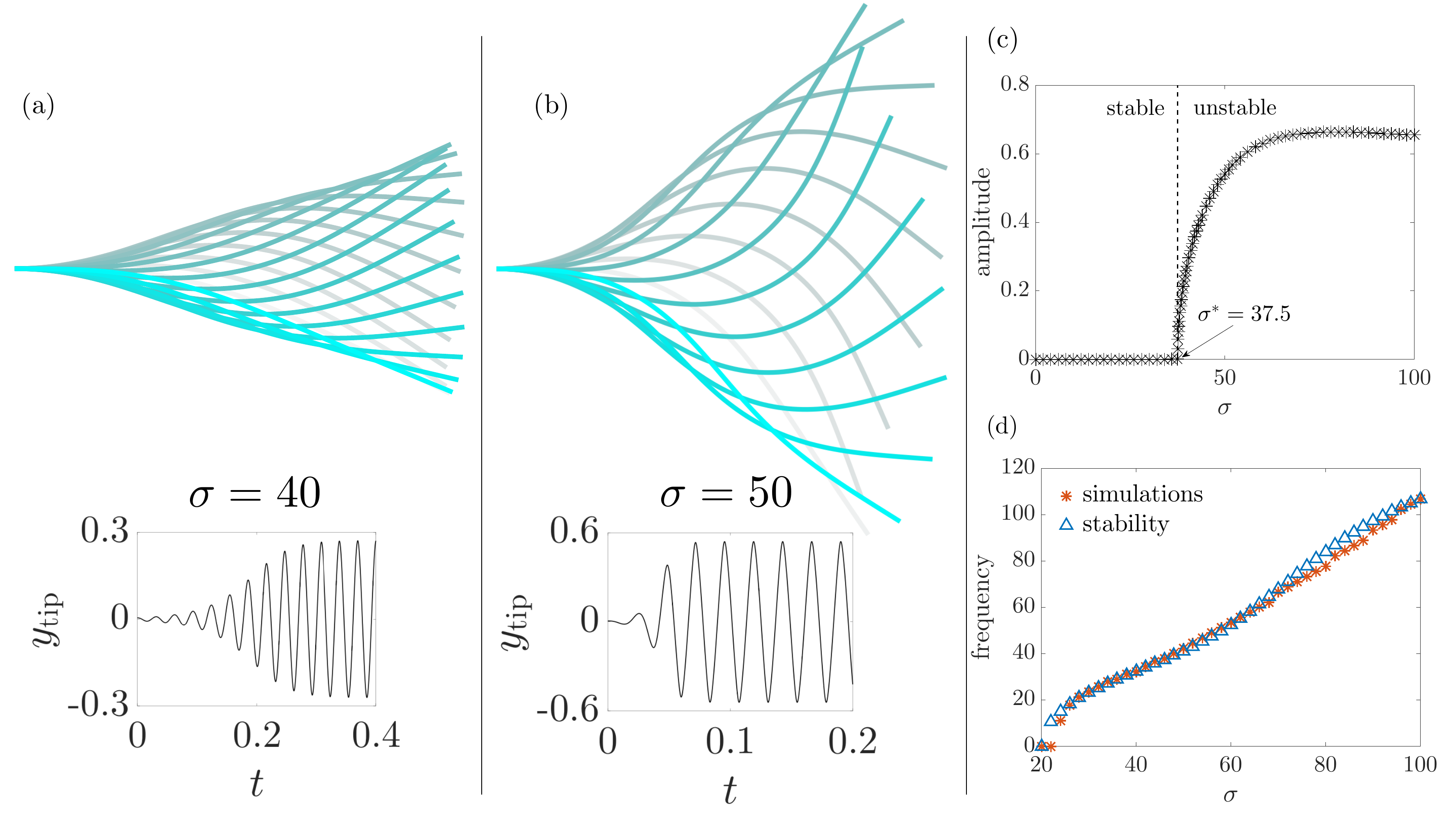}
\caption{Time evolution of flapping filament  and  of the tip displacement (inset) for different  values of $\sigma$ as obtained numerically; (a): $\sigma = 40$, (b): $\sigma = 50$. The transient required to  reach the 
oscillations decreases while the amplitude of the oscillations increases with $\sigma$. 
(c): Amplitude of the tip displacement as function of $\sigma$.  At $\sigma \approx 37.5$, the system 
becomes unstable and exhibits self-sustained oscillations.    (d): Comparison between the frequency of oscillation of the filament for different values of $\sigma$ 
obtained from the numerical simulations (red stars) and linear stability analysis (blue triangles).}
\label{misc}
\end{figure*}

Inspecting in more detail the dynamics of the filament  for $\sigma \gtrsim 37.5$ as shown in Fig.~\ref{misc}, we see that after a transient whose duration diminishes as  the value of $\sigma$ increases (Figs.~\ref{misc}a-b), the filament  traces a self-sustained wave  reminiscent of the waving of spermatozoa flagella \cite{brennen1977}.    The filament buckles as the external force keeps compressing it in the tangential direction
while both the elastic restorative force and the drag force oppose it,  giving rise to this {\it flapping
dynamics}.  It is worth stressing that this novel dynamics arises from the presence of the fluid in the low Reynolds number regime.  For an inertial filament with no fluid, the dynamics is indeed different 
\cite{bigoni2012}.
 We next plot in Fig.~\ref{misc}c  the amplitude of the 
oscillations   as a function of $\sigma$.  The tip displacement shows a clear Hopf bifurcation before  reaching a plateau 
(a consequence of the finite length of the filament).  The frequency of oscillation, 
which was computed applying the FFT to the time evolution of the tip displacement, grows roughly
linearly with $\sigma$ (Fig.~\ref{misc}d).  

\section{Linear Stability Analysis}
\label{linstab_sec}

The numerical results in the previous section reveal that  that increasing values of $\sigma$ are accompanied by a transition from 
stability to decaying oscillations, and finally a Hopf bifurcation to flapping dynamics.
We now turn to a theoretical analysis of this transition.

In order to study buckling instabilities, linear stability analysis has been exploited in several contexts, 
spanning from column buckling under compression -- a variant of Euler buckling-- with different 
boundary conditions (e.g. clamped-free, hinged-free, hinged-hinged, clamped-clamped) \cite{landau1986, 
bigoni2012}, 
to filament buckling in linear shear flow \cite{becker2001instability} or extensional flows 
\cite{young2007stretch, kantsler2012fluctuations, guglielmini2012buckling, deng2015}.  Because 
the follower force compresses the filament, a certain critical value above which the filament buckles is
expected to exist. Here, linear stability analysis is used to analytically compute the critical compression force. 

Assuming small deviations from the initial, straight configuration, Eq.~(\ref{forcebal}) simplifies 
as $x\approx s$, $\t \approx (1, y_x)$, and $\n \approx (y_x,-1)$.  The problem then turns into
solving the two coupled nonlinear equations (\ref{theta}) and (\ref{lambda}) to $\Lambda_x =0$, with 
$\Lambda (1,t) = \sigma$, which leads to $\Lambda(x,t)=\sigma$ and 
\begin{equation}
y_t = - y_{xxxx} - \Lambda y_{xx} \, ,
\label{linearised}
\end{equation}
with the boundary conditions 
\begin{equation}\label{eq:bc}
y(0,t) = y_x (0,t)= y_{xx}(1,t)=y_{xxx}(1) =0 \, .
\end{equation}

We first note that standard, so-called static methods {\it \`a la} Euler \cite{euler1774methodus}, fail to
predict buckling in our case, as consistent with classical analyses in the high-Reynolds number limit
\cite{timoshenko1970, bigoni2012}. A  static eigenvalue-based linear stability will only success in the  case 
where the forcing arises from conservative forces. In the situation considered in this paper, the external force
acts in a manner which depends  on the position and configuration of the entire filament, and is thus non-
conservative.  For systems with nonconservative forces in inertia-dominated problems, the critical value for which the
beam buckles and becomes unstable has been computed using a dynamic criterion 
\cite{beck1952knicklast,timoshenko1970}. Here, we   extend the analysis to 
the viscous-dominated regime.

We start by assuming a solution to the linearised problem,  Eq.~(\ref{linearised}), of the form 
\begin{equation}
y(x,t) = \hat y(x) \, \e^{\omega t} \, ,
\label{decomp}
\end{equation}
where $\omega$ is the growth rate.  This leads to the ordinary differential equation (ODE)
\begin{equation}
\hat y_{xxxx} + \sigma \hat y_{xx} + \omega \hat y = 0 \, ,
\label{eigenval_1}
\end{equation}
whose general solution  is given by
\begin{align}
\hat y(x) &=C_1 \cosh{\alpha_1 x} + C_2 \sinh{\alpha_1 x} + C_3 \cos{\alpha_2 x} + C_4 \sin{\alpha_2 x} \, ,
\end{align}
with 
\begin{align}
\alpha_1 & = \sqrt{\sqrt{\frac{\sigma^2}{4}-\omega} - \frac{\sigma}{2}} \, , \\
\alpha_2 & = \sqrt{\sqrt{\frac{\sigma^2}{4}-\omega} + \frac{\sigma}{2}} \, . 
\label{alpha_define}
\end{align}
The values of the constants $C_j$ are obtained by imposing the 
boundary conditions
in Eq.~\eqref{eq:bc}, leading to a standard $4\times 4$ determinental condition.  After some simplifications,
the equation for the growth rate, $\omega$, can be shown to read 
\begin{align}
\sigma^2 &- 2\omega\left(1+\cosh{\alpha_1}\cos{\alpha_2}\right)  + \sigma \sqrt{-\omega} \sinh{\alpha_1}\sin{\alpha_2} = 0 \, ,
\label{analytical_freq}
\end{align}
which, with the $\alpha_i$ defined in \eqref{alpha_define}, does not have a closed-form solution, 
but can be easily solved numerically.

Alternatively, we can also solve Eq.~(\ref{eigenval_1})  directly numerically, viewed either    as a boundary value 
problem or as an eigenvalue problem.  For the former, a shooting method is used with the appropriate initial guess in the neighbourhood of the first transition 
($\sigma = 20.05$). For the latter,  
the problem turns into solving $\mathcal{L} \hat y = \omega \hat y$. The operator $\mathcal{L}\equiv - 
\d^4/\d x^4 - \sigma \d^2/\d x^2$ is discretised using centred finite differences in the 
bulk of the stencil and sided differences at the ends, and the eigenvalues are determined with the QR algorithm.  Both  
methods were implemented and used to test the results obtained from the numerical solution of Eq. 
(\ref{analytical_freq}), showing excellent agreement.

\begin{figure}[t!]
	\centering
	\includegraphics[width=0.65\textwidth]{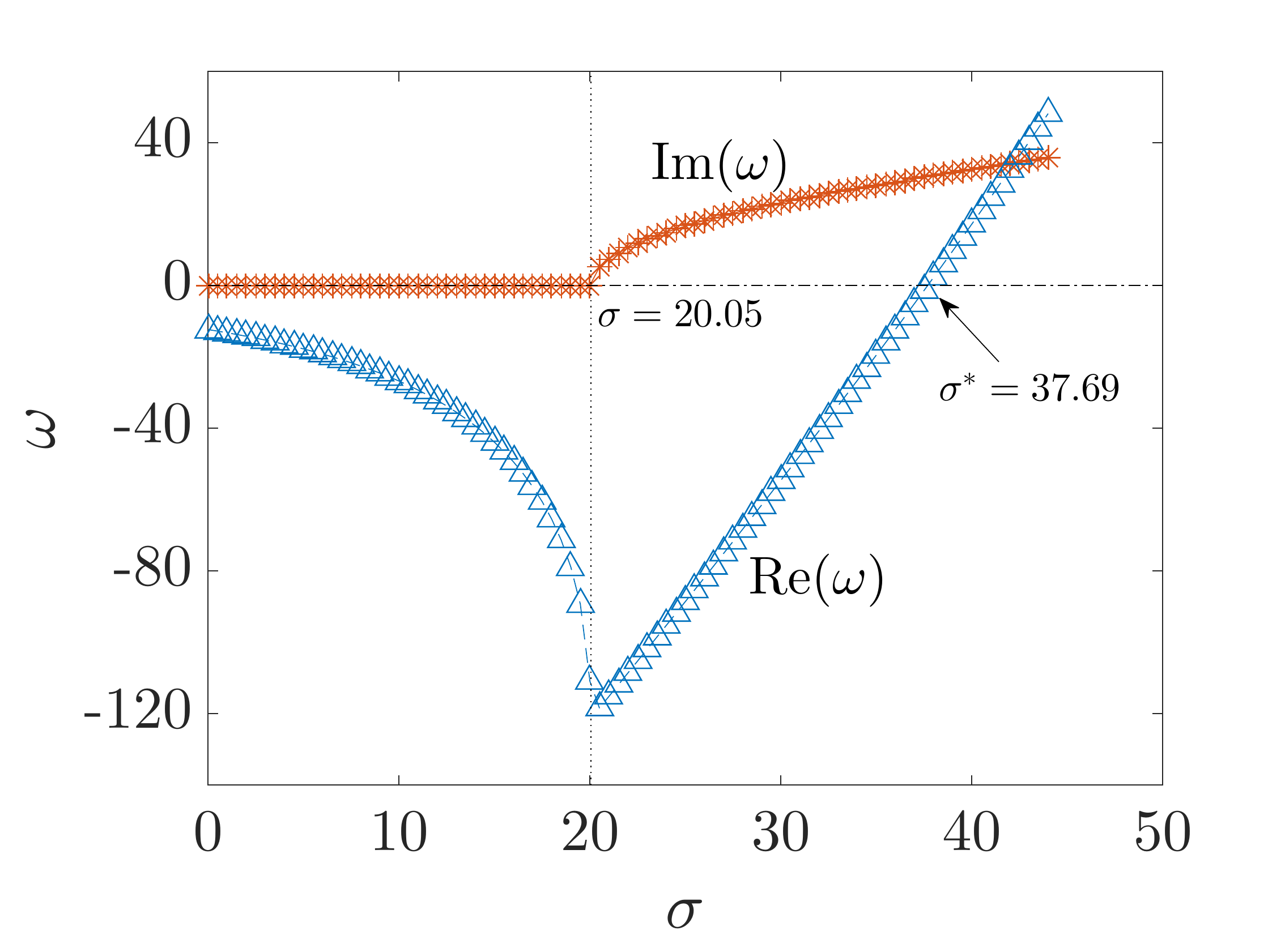}
	\caption{Imaginary and real part of the growth rate of the perturbation, $\omega$, as function of 
	$\sigma$. The frequency 
	becomes complex at $\sigma = 20.05$,  giving rise to oscillations in the filament dynamics. The real 
	part remains negative (stability) until $\sigma\approx  37.69$, after which it becomes positive (instability).}
	\label{transition_linear}
	\end{figure}

The linear stability results  identify three different behaviors as a function of the value of $\sigma$. These are illustrated in Fig.~\ref{transition_linear} where we plot the real part (blue triangles) and imaginary part (red stars) of the computed growth rate, $\omega$.  When 
$\sigma \lesssim 20.05$, the growth rate  is negative and $y(x,t)$ decays exponentially. Starting at $\sigma \gtrsim 20.05$, the growth rate becomes complex, but its real part remains negative,
as consistent with the numerical results  from the previous section  showing oscillatory decay. The real part of the growth rate finally becomes positive at a critical value, $\sigma^* \approx 37.69$ (Fig.~\ref{transition_linear}), indicating the onset of  the instability and the  bifurcation to oscillations about the horizontal, straight configuration. 

The comparison between the numerical results and linear stability analysis shows a very good 
agreement not only for the critical value of $\sigma$ at which the oscillations arise ($\sigma =20.4$ vs.~20.05) and at 
which the system becomes 
unstable ($\sigma^* =37.5$ vs.~37.69), but also for the frequency of oscillations (see Fig.~\ref{misc}b). Notably, 
the frequencies are also in good agreement also for large values of $\sigma$ when linear stability analysis 
does not strictly apply. 


\section{Two-link filament model}
\label{two-link_sec}
\begin{figure}[t]
  \centering
  \includegraphics[width=0.5\textwidth]{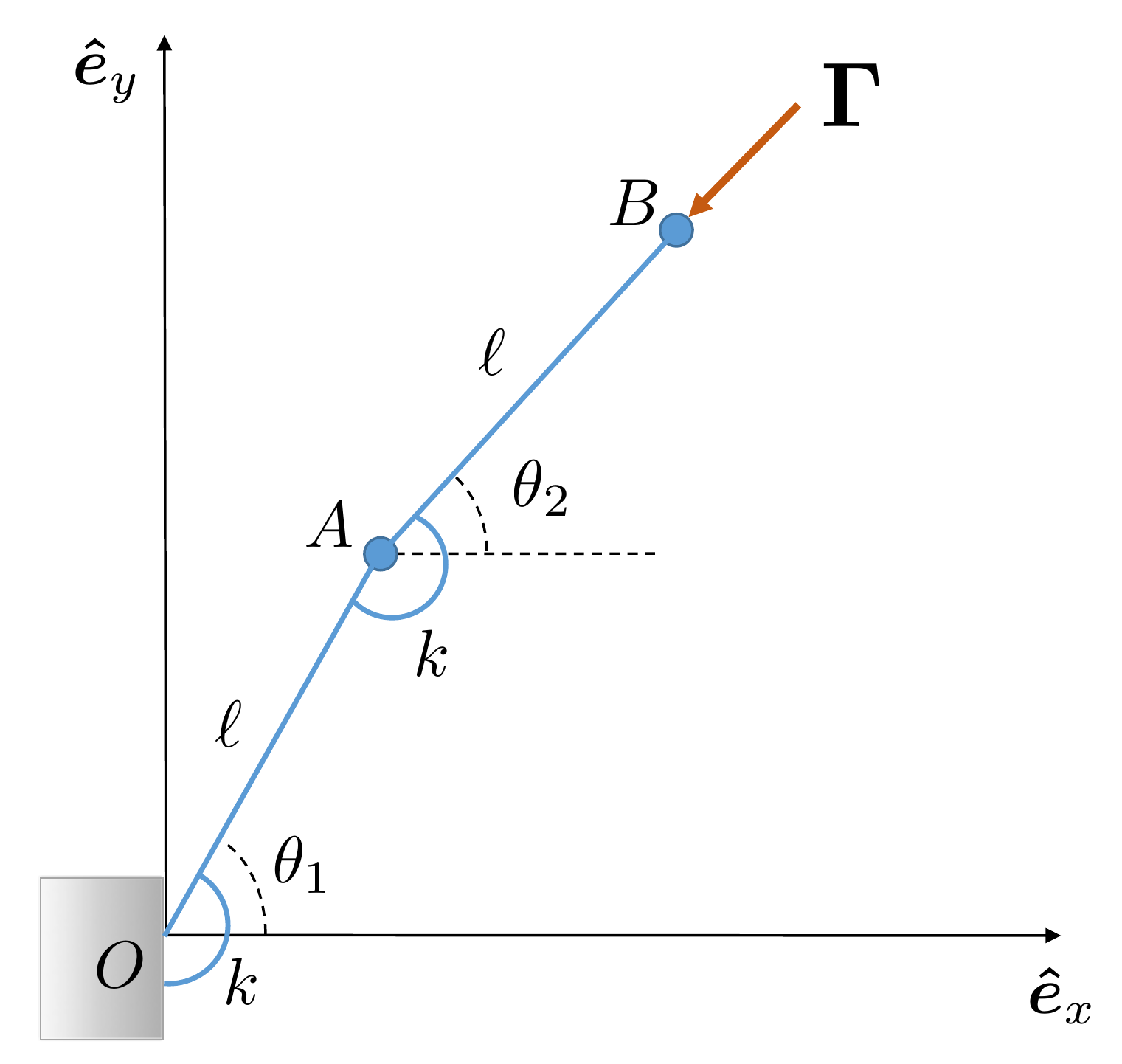}
  \caption{Discrete model: Two links  of length $\ell$ 
  rotate with  
  degrees of freedom $\theta_1$ and $\theta_2$ around  torsional  springs of strength 
    $k$ and are acted upon by
   a follower force  $\Gammavec$.}
  \label{fig5}
\end{figure}

Having shown that the linear stability analysis of the elastohydrodynamic PDEs can explain the onset of  flapping dynamics, we now consider a simpler   two-link filament model, in a manner   similar to the case in which damping is negligible \cite{ziegler1952, Herrmann1964}, with the aim of illustrating in a low-dimensional dynamical system the origin of oscillatory motion.

 We consider a simple discrete model for an elastic filament composed of two rigid links of length $\ell$ joined together 
at point $A$ and constrained to remain in the plane $z=0$ (see Fig.~\ref{fig5}). Elasticity is included by 
introducing two torsional springs, each with spring constant $k$.  The two degrees of freedom of the system are the angles $\theta_1(t)$ and $\theta_2(t)$ that define the configuration of the links.  They are  zero when both rods are horizontal and increase in the clockwise direction. The follower force,  $\Gammavec$, acts at the tip of 
the second rod, always pointing tangentially along it. The filament moves in a creeping flow and its drag
force is assumed to be concentrated at points $A$ and $B$ only.

For this model, the locations of points $A$ and $B$ are
\begin{align}
\r_A & = A-O = \ell \, (\cos\theta_1, \sin\theta_1) \, , \\
\r_B & = B-O = \ell \, (\cos\theta_1+\cos \theta_2, \sin\theta_1+\sin\theta_2) \, ,
\end{align}
and their velocities are
\begin{align}
\v_A  =\dot \r_A = & \ell  \, \dot \theta_1 (-\sin\theta_1, \cos\theta_1) \, , \\
\v_B  = \dot \r_B = & \ell  \, [ \dot \theta_1 (-\sin\theta_1, \cos\theta_1)  + 
												\dot \theta_2 (-\sin \theta_2, \cos\theta_2)] \, ,
\end{align}
where the dot denotes a time derivative. 
The follower force is defined as $\Gammavec = -\Gamma \t$, with $\Gamma >0 $ its magnitude and 
$\t = (\cos \theta_2, \sin \theta_2)$ the unit tangent vector joining $A$ and $B$. Under the assumption 
of creeping flow, the drag forces are $\F_A = -\zeta \v_A$ and 
$\F_B = -\zeta \v_B$, with $\zeta$ some effective drag coefficient, while the restoring moments 
due to the torsion springs 
acting on the two rods are $-k \theta_1$ at point $O$  and $-k (\theta_2 -\theta_1)$ at point $A$.

The equations of motion are obtained applying the principle of virtual work
\begin{equation}
\begin{split}
\Gammavec \cdot \delta \r_B +  \F_B \cdot \delta \r_B + \F_A \cdot \delta \r_A  - k \theta_1 \delta \theta_1 - k (\theta_2 -\theta_1) (\delta \theta_2 - \delta \theta_1) = 0 \, ,
\end{split}
\end{equation}
where $\delta \r_B, \delta \r_A, \delta \theta_1$ and $\delta \theta_2$ are the virtual displacements. 
Invoking the arbitrariness of $\delta \theta_1$ and $\delta \theta_2$, we obtain
\begin{eqnarray}
\Sigma \sin (\theta_1- \theta_2) -  [2 \dot \theta_1 +  \dot \theta_2\cos(\theta_1 - \theta_2)]  - 2 \theta_1 + \theta_2 &=& 0 \, , \label{two-link_1} \\
 - \dot \theta_1 \cos(\theta_1 - \theta_2)  \dot \theta_2  + \theta_1 - \theta_2&=& 0 \,, \label{two-link_2}
\end{eqnarray}
where time was rescaled by $\tilde t = kt/\zeta\ell^2$,
and we introduced the controlling dimensionless number, $\Sigma=\Gamma \ell/k$, playing a role 
similar to $\sigma$ in the previous section.   
Note that if we enforce $\theta_1 = \theta_2 = \theta \,$, then the previous equations reduce to
\begin{equation}
3\dot \theta + \theta = 0 \, ,
\end{equation}
which shows that the follower force, which always points inward, does not play any role and that 
$\theta$ decays exponentially, as we would expect.

We solved Eqs.~(\ref{two-link_1})-(\ref{two-link_2})
numerically  using the Matlab ODE solver `ode45', which is based on an explicit Runge-Kutta 
(4,5) formula and is suitable in this case as the equations are non-stiff \cite{dormand1980}.  The initial 
conditions are random, small perturbations to both angles.

\begin{figure*}[t!]
	
	\includegraphics[width=\textwidth]{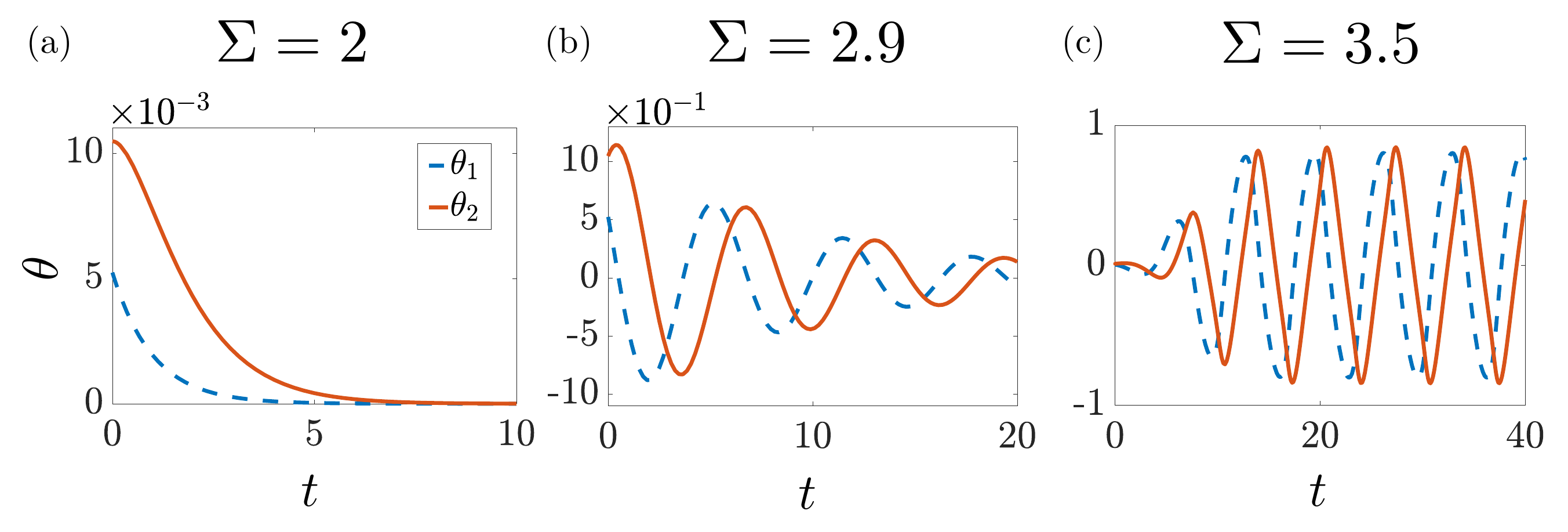}
	\vspace*{-15pt}
	\caption[]{Time evolution of $\theta_1$ (blue dashed line) and $\theta_2$ (red solid line)
	after solving numerically the nonlinear equations of motion. 
	(a): The system is asymptotically stable 
	($\Sigma=2$); 
	(b): Stability with  oscillations  ($\Sigma=2.9$); 
	(c): Stable, self-sustained oscillations
	($\Sigma=3.5$).}
	\label{two_link_full}
\end{figure*}

Our numerical results, 
shown in Fig.~\ref{two_link_full},  
indicate that, again, three 
different dynamics are possible. With increasing values of $\Sigma$, the system goes from  
asymptotic stability ($\Sigma<2$), to stability with 
oscillations  ($2\le\Sigma<3$),  to exhibiting stable, self-sustained oscillations ($\Sigma\ge 3$).

In order to  capture these transitions, we may again take advantage of linear stability. 
By linearising the equations of motion about the equilibrium configuration $\theta_1 = \theta_2 = 0$, 
and assuming solutions of the form $\theta_j = \hat \theta_j \, \e^{\omega \tilde t}$ we 
obtain 
\begin{align}
&\Sigma  (\hat \theta_1- \hat \theta_2) - \omega (2 \hat \theta_1 + \hat \theta_2) - 2 \hat \theta_1 +  \hat \theta_2 = 0, \\
& - \omega (\hat \theta_1 + \hat \theta_2) +
		\hat \theta_1 - \hat \theta_2 = 0 \, ,
\end{align}
and non-trivial solutions are found when the determinant of the corresponding matrix is zero, namely
$\omega^2 + 2 (3 - \Sigma) \omega + 1 = 0$,
whose solutions are
\begin{equation}
\omega_{\pm} =  \Sigma - 3\pm \sqrt{\left( \Sigma-4\right) \left( \Sigma-2\right)} \, .
\label{sol_omega}
\end{equation}

We may then use Eq.~\eqref{sol_omega} to predict the dynamics, and we obtain five different 
cases: 
\begin{enumerate}
\item[a.] if $\Sigma \le 2$, then $\omega_{\pm} <0$, and the system is stable; 
\item[b.] for $ 2<\Sigma< 3$, $\rm{Re}(\omega_{\pm}) <0$ and $ \rm{Im}(\omega_{\pm}) \ne 0$,
so the perturbations die away in an oscillatory manner, 
\item[c.]if $\Sigma = 3 $, then 
$\rm{Re}(\omega_{\pm}) =0$ and $ \rm{Im}(\omega_{\pm}) \ne 0$, hence the system is stable and 
shows periodic oscillations with constant amplitude; 
\item[d.] for $3 < \Sigma < 4$, 
$\rm{Re}(\omega_{\pm}) >0$ and $ \rm{Im}(\omega_{\pm}) \ne 0$, and thus we obtain  
exponentially-growing oscillations; 
\item[e.]when $\Sigma \ge 4$, $\omega_{\pm} >0$, i.e. the system is unstable 
and $\theta_1,\theta_2$ simply diverge.  
\end{enumerate}
In cases d-e, the linear instability saturates to nonlinear self-sustained oscillations when the full nonlinear equation is considered. Once again, linear stability is thus in  good agreement with the results from the nonlinear equations
of motion.

In conclusion, the two-link model studied in this section captures the dynamics of the full
nonlinear elastohydrodynamic problem.  In particular, we have shown that when 
$\Sigma = \Gamma \ell / k \ge 3$, which represents, analogously to $\sigma$, the ratio between 
the strength of the follower force and the elastic force, self-sustained oscillations are indeed
possible.

\section{Physical interpretation}

The analysis in Sec. \ref{insights} showed that the boundary term of the RHS of Eq.~(\ref{E_evolve}) 
arises from the nonvariational nature of the follower force.  
Here we simulate the full nonlinear elastohydrodynamics equations 
and demonstrate that it is indeed the term responsible for the self-sustained motion observed.

\begin{figure*}[t]
\centering
   	\includegraphics[width=\textwidth]{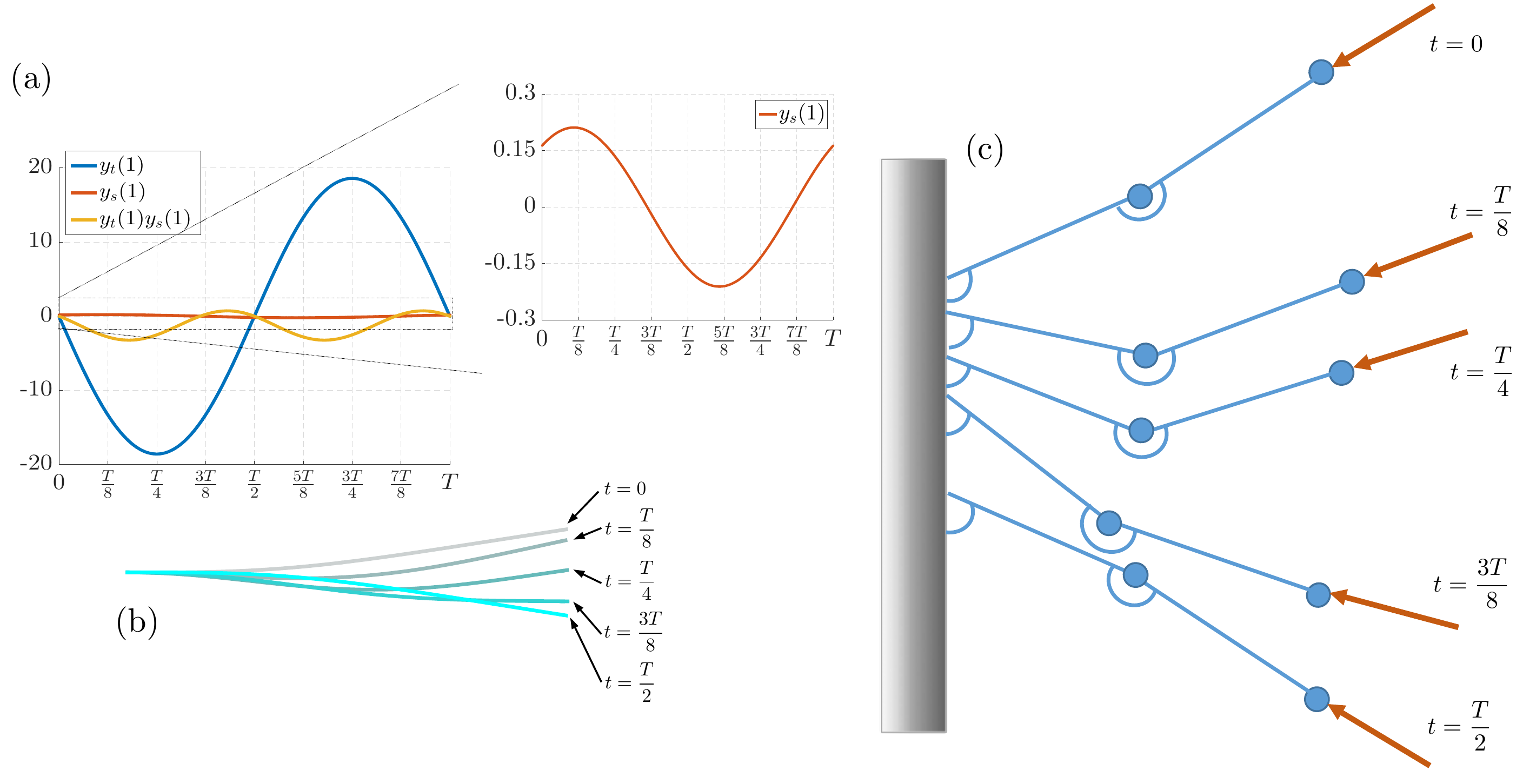}
	\vspace*{-15pt}
	\caption[]{Nonvariational aspects of flapping motion.  (a) Time evolution of the tip velocity, 
$y_t(1)$, its derivative with respect to the arclength, $y_s(1)$, (zoomed in the inset) 
and their product, $y_t (1) \, y_s(1)$ 
over the period of oscillation $T$ for $\sigma = 37.8$.  The term $y_t (1) \, y_s(1)$ 
changes sign four times over a cycle: the presence of the follower force both removes and injects 
energy into the system, giving rise to self-sustained, periodic oscillations.  (b) Filament configuration 
over a half-period.  (c) Schematic of the two-link model at different times over a half-period for 
$\Sigma = 37.5$.  The follower 
force compresses the two-link structure and the links are pushed 
downward ($0<t<T/4$). Then, the first link reaches its lowest point (i.e. highest restorative moment) and
stops moving, while the second link keeps rotating ($t=3T/8$).  By doing so, the follower force, which 
has followed the second link, exerts a lower moment and the restorative effect becomes predominant. 
Hence, the first link moves upwards and the second link downwards until stopping and inverting its motion 
($t=T/2$).  The cycle then repeats.}
\label{physics}
\end{figure*}

Choosing the value $\sigma=37.8$ allows the tip oscillations to remain small.  We plot in Fig.~\ref{physics}a 
the values of the tip velocity, $y_t(1)$, slope, $y_s(1)$, and their product.   Over the period of oscillation
$T$, which is defined such that the tip displacement is maximum at $t=0$, the tip reaches the minimum at 
$t=T/2$ and crosses the $x$-axis twice, with minimum and maximum speeds 
at $t=T/4$ and 
$t=3T/4$, respectively.  In contrast, the filament tangent at the tip, $y_s(1)$, has its maximum value at about $t=T/8$ 
and minimum at about $t=5T/8$, becoming zero slightly before $t=3T/8$ and $t=7T/8$.  

While the term $-\sigma y_t (1) \, y_{s}(1)$ is positive, it injects energy into the system until 
the tangent at the tip crosses the $x$-axis.  At this point, it becomes negative and it therefore
withdraws 
energy until the tip reaches its minimum displacement.  Afterwards, it becomes
positive again and the cycle repeats, but with the mirrored configuration ($T/2<t<T$).  For reference,
we show in Fig.~\ref{physics}b  the filament configuration over a half-period.

In order to  better understand this dynamics, we may also exploit the two-link model previously studied, with dynamics illustrated Fig.  
\ref{physics}c for $\Sigma=3$.  Initially, the follower force compresses the two-link structure and the links are pushed 
downward ($0<t<T/4$). Then, the first link reaches its lowest point (i.e. highest restorative moment) and
stops moving, while the second link keeps rotating ($t=3T/8$).  By doing so, the follower force, which 
has followed the second link, exerts a lower moment and the restorative effect becomes predominant. 
Hence, the first link moves upwards and the second link downwards until stopping and inverting its motion 
($t=T/2$).  This dynamics repeat periodically and prevents the establishment of a steady state.

By examining both the continuous and discrete models, we thus see that the 
 effect of the follower force is to constantly inject and remove energy into the system, thus preventing any stable configuration to be reached and giving rise to periodic, 
self-sustained oscillations.


\section{Flow-entraining follower force}
\label{flow_sec}

\begin{figure}[t]
  \centering
  \includegraphics[width=0.8\textwidth]{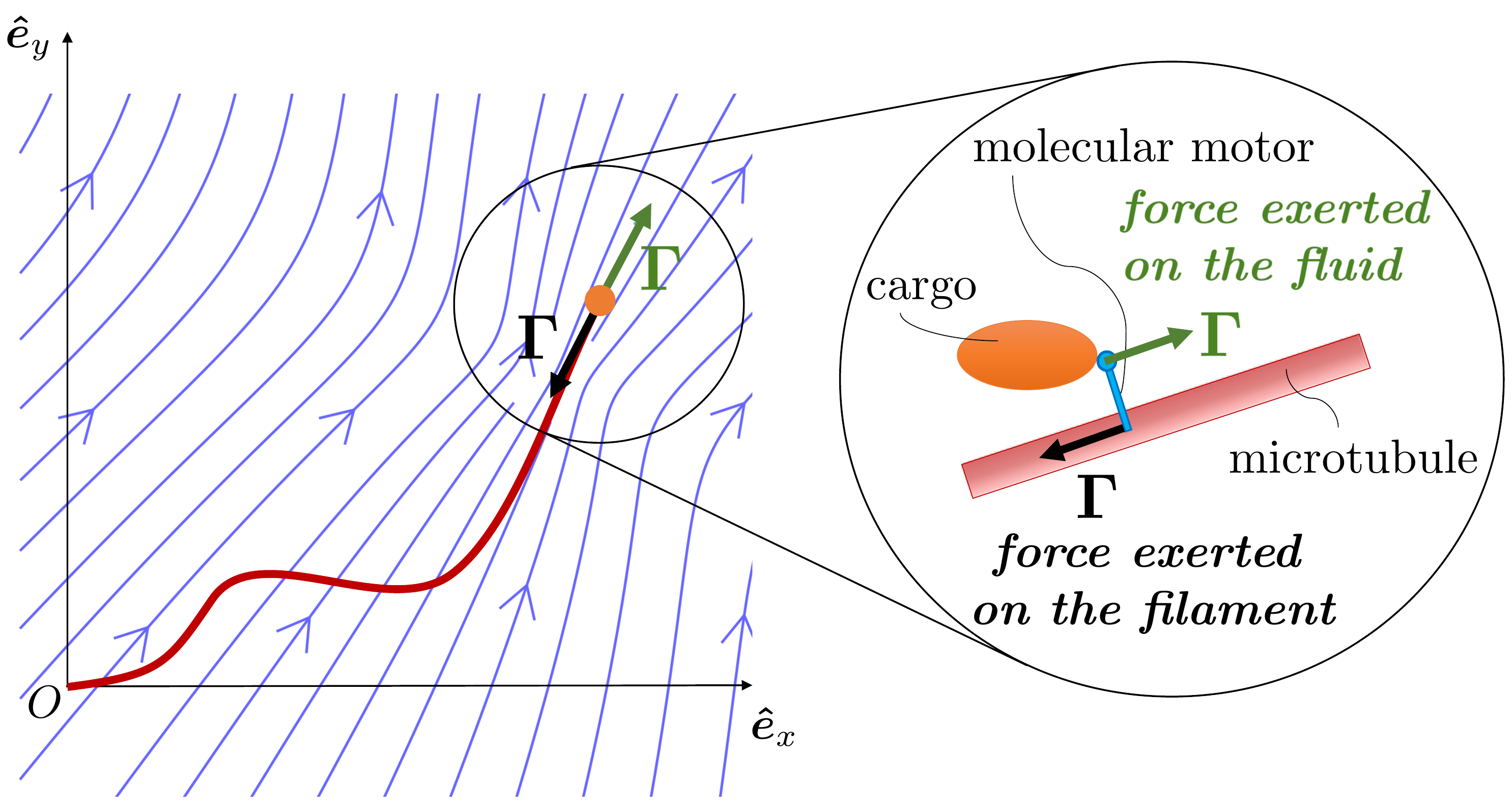}
  \caption{A filament clamped on one side ($s=0$) and subject to a `flow-entraining' follower force at the other end ($s=L$).  
  At the tip,  the compressive follower force acts on the filament (black arrow) while the moving cargo acts on the 
  surrounding fluid as a  point force (green) creating the flow with streamlines illustrated in blue.  Inset: A detailed picture 
  of the forces acting on the filament and the fluid.}
  \label{activegeom}
\end{figure}

When molecular motors translocate along MTs they not only exert a force 
on the filament, but also entrain fluid as they carry cargo.  While the motor-induced force on the 
filament was included in the analysis of previous sections, the fluid flow created by the motor and the 
associated drag on the filament were neglected. 
Here we include these effects by approximating cargo-motor assembly as a point force located at the tip of the filament, so that the three-dimensional (3D) flow that arises is that of a stokeslet \cite{chwang75}.  As at the tip we now have both a concentrated load acting on the filament
(the follower force) and a concentrated force setting the flow, we refer to this combination as 
`flow-entraining' follower force (see illustration in Fig.~\ref{activegeom}).

\subsection{Equations of motion}
When molecular motors walk along the MT towards its free end, they create a flow that follows their 
direction of motion while applying a force on the filament in the opposite direction.  We assume that
the link between the filament and the cargo to be rigid.  As the magnitude of the force exerted 
on the filament while the molecular motor walks along it is $\Gamma$, a simple force balance shows 
that the force exerted on the fluid also has strength $\Gamma$ (inset of Fig.~\ref{activegeom}).
The fluid flow on the filament centerline created by the point force at $s=L$ 
is therefore $\u (s) = (1/8\pi\mu)  \G(s;L)\cdot (\Gamma \t(L))$, where $\G(s;L)$ is the 
Green's function tensor (with dimensions of inverse length) appropriate to the 
boundary conditions imposed on the fluid equations.   In the following, we consider 
the 3D fluid flow created by a point force in an unbounded domain, and thus ignore the presence of any boundary (though the analysis could be repeated in this case along the same lines).  

The drag force acting on the filament using the RFT 
approximation is now given by 
\begin{equation}
\f_h = - \left( \zpar \t \t + \zper \n \n \right) \cdot \left( \r_t - \u \right),
\end{equation}
and thus the equations of motion become
\begin{equation}
 \left( \zpar \t \t + \zper \n \n \right) \cdot \left( \r_t - \u \right) = 
- A \r_{ssss} - \left( \Lambda \r_s \right)_s \, ,
\label{forcebal_active}
\end{equation}
or, in dimensionless form,
\begin{equation}
\begin{split}
 \left(\eta^{-1} \t \t + \n \n \right) \cdot \left( \r_t - \xi \sigma \u \right) = 
- \r_{ssss} - \left(\Lambda \r_{s}\right)_s \, ,
\label{forcebal_active_nd}
\end{split}
\end{equation}
where $\xi=(\zper/8\pi\mu)$.  
Here we assume that the flow only comes from the point force, and neglect 
the fluid flow that arises from the filament's motion.  This effect will be discussed elsewhere \cite{decanio2017inprep}.

The Green's function is singular at $s=L$, thus a regularization is needed in order to avoid 
overestimating the magnitude of the velocity produced by the point force.  In order to achieve this, we use the expression 
for a regularized Stokeslet derived by Cortez et al. \cite{cortez2008}, characterized by a single
regularization parameter $\delta$. 

In order to set the value of  $\delta$, we use the following physical argument.  The 
velocity field at a distance $r$ from a regularized point force with strength $F$ decays as 
$u\sim F/8 \pi \mu 
(r + \delta)$.  We require that the magnitude
of the fluid flow at the location of the point force, $F/8\pi\mu\delta$  be equal to the motor speed 
$u_{\rm motor}$.  In order to determine $\delta$ we thus need to know the  
magnitude of the point force, the speed of the molecular motor, and the viscosity of the medium.
Our work was inspired by phenomena involving cytoplasmic streaming in {\it Drosophila} oogenesis, where
the measured viscosity can reach $\mu \approx 1$
Pa s \cite{Ganguly2012}, three orders of magnitude larger than water.   Typical molecular motors speeds 
in animals are fractions of microns/sec, while the forces they exert are on the piconewton scale \cite{svoboda1994}.
Considering the full range of viscosities we obtain $\delta \approx 10^{-7}-10^{-4}$ m, the smaller values associated 
with the higher viscosities.  Adopting the value $10^{-6}$ m as representative of the situation in 
\textit{Drosophila}, we see that $\delta/L \sim 0.05-0.1$ as MTs are usually some $10-20$ $\mu$m 
long \cite{Ganguly2012}.

Let $\tilde \u =\t(1) \cdot \tilde \G(s;1)$, with $\tilde \G(s;1)$ the regularized Green's tensor
\cite{cortez2008}.  The generalization of Eqs.~\eqref{theta} and \eqref{lambda} to the flow-entraining 
force is 
\begin{eqnarray}
\theta_t  &=& 	-\theta_{ssss} 
			- \left[ \Lambda - 3 \left(\eta +1 \right) \theta_s^2 \right]\, \theta_{ss}     - \left (\eta +1 \right) \Lambda_s \, \theta_s  
			- \xi \, \sigma \, \tilde \u_s \cdot \n \, , 
\label{theta_active} \\
\Lambda_{ss} - \eta^{-1}\theta_s^2 \Lambda  &=& 
			-\eta^{-1} \theta_s^4 
			+ 3 \theta_{ss}^2  
			+ \left( 3+\eta^{-1} \right) \theta_s \, \theta_{sss}  + \eta^{-1}\xi \, \sigma \, \tilde \u_s \cdot \t \, . 
\label{lambda_active}
\end{eqnarray}
While the boundary conditions at the free end remain the same, an evaluation of  Eq.
(\ref{forcebal_active_nd}) at $s=0$ shows that the presence of the background flow leads to the condition
\begin{equation}
\theta_{sss}(0,t) - \theta_s(0,t)^3 + \theta_s(0,t) \Lambda(0,t) +
\xi \, \sigma \, \tilde \u(0,t)\cdot \n(0,t) = 0,
\end{equation}
for the tangent angle and 
\begin{equation}
\Lambda_s(0,t) - 3 \, \theta_s (0,t) \, \theta_{ss} (0,t) - \eta^{-1} \xi \, \sigma 
\tilde \u(0,t) \cdot \t(0,t) = 0 \end{equation}
 for the Lagrangian multiplier.    Note that while in this study we ignore the presence of any boundaries from a hydrodynamic standpoint, the value of $\tilde \u(0,t)$ would be set to zero if the Green's function used was the one which includes the  presence of the wall \cite{blake74_image}.

\subsection{Linear stability analysis}
By projecting Eq.~(\ref{forcebal_active_nd}) in the normal and tangent directions and 
after neglecting higher order terms, we obtain 
\begin{equation}
\Lambda_x = \eta^{-1}\xi \sigma \tilde u,
\end{equation}
and 
\begin{equation}
y_t = -y_{xxxx}  - \Lambda y_{xx} + \xi \sigma ( \tilde v - \tilde u \, y_x ),
\label{LSA_active_y}
\end{equation}
where $\tilde u = K_1 + K_2 (x-1)^2$ and $\tilde v = K_1 y_x(1) + K_2 (x-1)[y-y(1)]$ are the 
linearized components of the regularized nondimensional fluid flow $\tilde \u$, 
with 
\begin{equation}
K_1 = \frac{(x-1)^2 + 2 \delta^2}{[(x-1)^2 + \delta^2]^{3/2}}, \ \ 
K_2 = \frac{1}{[(x-1)^2 + \delta^2]^{3/2}}\cdot
\end{equation} 
We note that far away from the point force ($\vert x-1\vert \gg \delta$) the dominant flow component falls off as
that of a stokeslet, $\tilde u\sim 1/ \vert x-1 \vert$.
Interestingly, the term $\Lambda_x \, y_x$ does not 
appear in Eq.~(\ref{LSA_active_y}) as the product $\n \cdot \Lambda_x \r_x$ is identically zero.  
In other words, the fact that the tension varies along the filament length enters the equation only 
through $\Lambda$, but not its derivative.  

We use the same dynamic criterion described previously to determine the value at which the 
filament buckles and becomes unstable.  We compute the Lagrange multiplier first, requiring
$\Lambda(1,t)= \sigma$, and then solve by finite differences the eigenvalue problem 
\begin{equation}
\begin{split}
\hat y_{xxxx} - & \Lambda \hat y_{xx} + \xi \sigma \{K_1 \hat y_x(1) + 
K_2 (x-1) [\hat y-\hat y(1)]  - \hat y_x [K_1 + K_2(x-1)^2] \} = \omega \hat y \, ,
\end{split}
\label{eigenval}
\end{equation}
with the boundary conditions $ \hat y(0)= \hat y_x(0)= \hat y_{xx}(1)= \hat y_{xxx}(1)=0$.  Through the dynamics of the tip, the  hydrodynamic point force 
changes  position in time, and thus the resulting fluid flow is time-dependent.   
This is the origin of the
terms containing $\hat y(1)$ and $\hat y_x(1)$ in \eqref{eigenval}.  The numerical implementation
of this eigenvalue problem is more challenging than in the absence of entrained flow and great care
is needed, especially when discretising the boundary conditions and the local terms.

\subsection{Results}

\begin{figure*}[t!]
\centering
   \includegraphics[width=\textwidth]{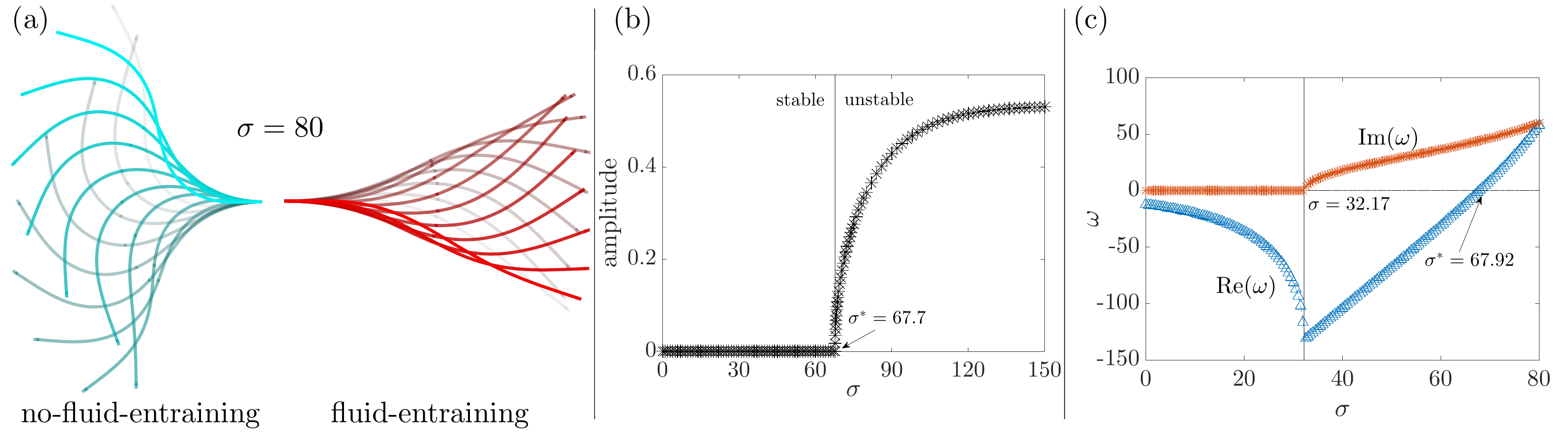}		
	\caption{(a) Time evolution of filament flapping and of the tip displacement (inset) for 
	the no-fluid-entrained follower force of Sec. \ref{buckling_sec} (blue) vs.~the fluid-entrained follower force (red).  The fluid flow reduces the
	tension on the filament,  resulting in a delay of the instability and  lower amplitude self-sustained oscillations.  (b) The system
	undergoes a supercritical Hopf bifurcation at $\sigma^* = 67.7$.  (c) Imaginary and 
	real parts of the growth rate, $\omega$, as function of $\sigma$.  The growth rate becomes complex 
	at $\sigma = 32.17$, thus giving rise to oscillations in the filament dynamics; its real 	part remains
	negative until $\sigma^* = 67.92$.  For larger values of $\sigma$, the real part becomes positive, hence 
	leading to instability.}
\label{active_comp}
\end{figure*}

The equations of motion were  
solved numerically using the procedure  described in   Sec. \ref{buckling_sec}.  Unsurprisingly, 
the dynamics has remained qualitatively unaltered, as shown in  Fig.~\ref{active_comp}a.  Here again three dynamical regimes may be identified. The 
filament starts showing decaying oscillations at $\sigma = 32.4$ and becomes unstable undergoing
a supercritical Hopf bifurcation at $\sigma^* = 67.7$ (Fig.~\ref{active_comp}b).   The transition points between the different regimes are well captured by linear analysis which predicts the growth rate to become
complex at $\sigma = 32.17$ (decaying oscillations) and to cross the imaginary axis
at $\sigma^* = 67.92$ (Hopf bifurcation).

Why is the flow delaying the onset of self-sustained oscillations? 
The point force located at the tip of the filaments induces a fluid flow in the  direction opposite to the
compressive force,  resulting in an added tension along the filament, and thus an effective compression 
which is lower than that of the no-flow-entraining follower force case.  Consequently, 
the transition from stable to unstable occurs at a larger value of $\sigma$.

\section{Discussion}
\label{discussion_sec}

Inspired by experimental observations of persistent waving motion of MTs driven by molecular
motors, particularly during oocyte development in {\it Drosophila} \cite{Ganguly2012}, we have explored
the simplest model of motor-driven filament motion.  In this ``follower-force'' model, a compressive
motor force ${\bf \Gamma}$ acts tangentially at the free end of the filament whose shape is found by 
balancing the forcing with elasticity and low-Reynolds number fluid drag.   Numerical studies of the full 
nonlinear elastohydrodynamics 
equations   led to the discovery of a flapping instability
that arises as the control parameter, $\sigma=\Gamma L^2/A$, is varied.  As is typically the case in a
Hopf bifurcation, the linearized filament dynamics first develops damped oscillations at an intermediate
value of $\sigma$ before exhibiting self-sustained limit cycle motion beyond some 
critical value, $\sigma^*$, both of which are also well captured by a linear stability analysis.  

Motivated by these findings,  we then proposed as a  toy model   a discrete two-link system in which
elasticity was  included via two torsion springs.  Linear stability analysis of this simpler dynamical system 
identified five 
different regions depending on the value of the control parameter $\Sigma = \Gamma \ell / k$, in full 
agreement with the results of numerical simulations.

Molecular motors entrain fluid while moving along microtubules.  In order to capture this effect, we next
developed a more realistic  continuum model based on approximating the forcing of the motor with its cargo 
on the surrounding fluid as that due to a localised force.   Since motors are known to walk towards the free 
end of microtubules, the flow they create point in the same direction, thus creating an effective flow-induced 
tension  and delaying the onset of flapping.  Although
the details of buckling are quantitatively different in the presence of this induced fluid flow, the physics of 
flapping is essentially the same. 

Having quantified the value for the onset of oscillations, it is important to relate it to the biological 
system which motivated its study, namely the {\it Drosophila} oocyte.  The force exerted on the filaments 
by the molecular motors is known to be, as already discussed, on the order of piconewtons and MTs are
approximately $20 \mu$m in length.  Despite the lack of information in the literature about the bending
modulus of MTs in this specific context, we may estimate their rigidity from the direct measurements 
by Gittes et al. for a single MT {\it in vitro}, $A\approx 10^{-23}$~N m${^2}$ \cite{gittes1993}. 
With these numbers, we obtain that $\sigma \approx 120$, indicating that the forcing from molecular 
motors is large enough to lead to buckling and oscillations in the biological system. 


The work in this study is but a first step towards capturing the full interplay of elastic and fluid mechanical 
forces  in cytoplasmic streaming. We have focused our analysis on the case of a single filament in an infinite 
fluid and subject to a force localised at its end. In order to capture biological dynamics, these simplifications 
should be relaxed,  in particular since we know that:  
(i) Multiple kinesin motors walk about each microtubule filament, possibly interacting hydrodynamically; 
(ii) Motors stochastically bind and unbind to the filaments, providing stochasticity to both the long-range 
forces in the fluid and the localised forces to the filaments; 
(iii) Microtubules are not found in isolation but tend to be densely packed, and therefore subject to   steric 
and hydrodynamic interactions; 
(iv) In the specific case of \textit{Drosophila} which motivated this study, the entire motor protein-filament 
network is located inside a closed cavity  (the oocyte), and the confinement of an incompressible fluid  
provides another way for filament to undergo long-range interactions. 
The  example of fluid-structure interaction addressed in this study will provide a fundamental basis to tackle 
these extensions and address the  dynamics of complex systems in  cellular biophysics.

\subsection{Acknowledgements}
This work was supported in part by ERC Advanced Investigator Grant 247333 (REG), an ERC 
Consolidator Grant 682754  (EL), and the Schlumberger Chair Fund.

\bibliographystyle{unsrt}
\bibliography{DeCanio_biblio}

\begin{thebibliography}{10}

\bibitem{Corti}
B.~Corti.
\newblock {\em Osservazione Microscopiche sulla Tremella e sulla Circulazione
  del Fluido in Una Planto Acquaguola}.
\newblock Appresso Giuseppe Rocchi, Lucca, Italy, 1774.

\bibitem{streaming_jrsi}
R.E. Goldstein and J.-W. van~de Meent.
\newblock A physical perspective on cytoplasmic streaming.
\newblock {\em Interface Focus}, 5:20150030, 2015.

\bibitem{Wasteneys}
I.~Foissner and G.O. Wasteneys.
\newblock Microtubule disassembly enhances reversible cytochalasin-dependent
  disruption of actin bundles in characean internodes.
\newblock {\em Protoplasma}, 214:33--44, 2000.

\bibitem{WoodhousePNAS}
F.G. Woodhouse and R.E. Goldstein.
\newblock Cytoplasmic streaming in plant cells emerges naturally by
  microfilament self-organization.
\newblock {\em Proc. Natl. Acad. Sci. USA}, 110:14132--14137, 2013.

\bibitem{oogenesis_ref}
W.E. Theurkauf, S.~Smiley, M.L. Wong, and B.M. Alberts.
\newblock Reorganization of the cytoskeleton during drosophila oogenesis:
  implications for axis specification and intercellular transport.
\newblock {\em Development}, 115 (4):923--936, 1992.

\bibitem{Ganguly2012}
S.~Ganguly, L.S. Williams, I.M. Palacios, and R.E. Goldstein.
\newblock Cytoplasmic streaming in drosophila oocytes varies with kinesin
  activity and correlates with the microtubule cytoskeleton architecture.
\newblock {\em Proc. Natl. Acad. Sci. USA}, 109:15109--15114, 2012.

\bibitem{Bourdieu}
L.~Bourdieu, T.~Duke, M.B. Elowitz, D.A. Winkelmann, S.~Leibler, and
  A.~Libchaber.
\newblock Spiral defects in motility assays: a measure of motor protein force.
\newblock {\em Phys. Rev. Lett.}, 75:176--179, 1995.

\bibitem{Gittes}
F.~Gittes, E.~Meyh{\"o}fer, S.~Baek, , and J.~Howard.
\newblock Directional loading of the kinesin motor molecule as it buckles a
  microtubule.
\newblock {\em Biophys. J.}, 70:418--429, 1996.

\bibitem{Young2010}
Y.-N. Young.
\newblock Dynamics of a semiflexible polar filament in stokes flow.
\newblock {\em Phys. Rev. E}, 82:016309, 2010.

\bibitem{IseleHolder}
R.E. Isele-Holder, J.~Elgeti, and G.~Gompper.
\newblock Self-propelled worm-like filaments: spontaneous spiral formation,
  structure, and dynamics.
\newblock {\em Soft Matter}, 11:7181--7190, 2015.

\bibitem{Chaudhuri}
A.~Chaudhuri and D.~Chaudhuri.
\newblock Forced desorption of semiflexible polymers, adsorbed and driven by
  molecular motors.
\newblock {\em Soft Matter}, 12:2157--2165, 2016.

\bibitem{Gosselin}
P.~Gosselin, H.~Mohrbach, I.~M. Kuli{\'c}, and F.~Ziebert.
\newblock On complex, curved trajectories in microtubule gliding.
\newblock {\em Physica D}, 318-319:105--111, 2016.

\bibitem{decanio2017inprep}
G.~De~Canio, R.E. Goldstein, and E.~Lauga.
\newblock {\em in preparation}, 2017.

\bibitem{Herrmann1964}
G.~Herrmann and R.~W. Bungay.
\newblock On the stability of elastic systems subjected to nonconservative
  forces.
\newblock {\em J. Appl. Mech.}, 31:435--440, 1964.

\bibitem{Langthjem}
M.~A. Langthjem and Y.~Sugiyama.
\newblock Dynamic stability of columns subjected to follower loads: a survey.
\newblock {\em J. Sound Vib.}, 238:809--851, 2000.

\bibitem{Elishakoff}
I.~Elishakoff.
\newblock Controversky associated with the so-called ``follower forces":
  critical overview.
\newblock {\em Appl. Mech. Rev.}, 58:117--142, 2005.

\bibitem{BaylyDutcher}
P.V. Bayly and S.K. Dutcher.
\newblock Steady dynein foces induce flutter instability and propagating waves
  in mathematical models of flagella.
\newblock {\em J. R. Soc. Interface}, 13:20160523, 2016.

\bibitem{audoly2010}
B.~Audoly and Y.~Pomeau.
\newblock {\em Elasticity and geometry: from hair curls to the non-linear
  response of shells}.
\newblock Oxford University Press, 2010.

\bibitem{goldstein1995nonlinear}
R.E. Goldstein and S.A. Langer.
\newblock Nonlinear dynamics of stiff polymers.
\newblock {\em Phys. Rev. Lett.}, 75(6):1094, 1995.

\bibitem{gray1955propulsion}
J.~Gray and G.J. Hancock.
\newblock The propulsion of sea-urchin spermatozoa.
\newblock {\em J. Exp. Biol.}, 32(4):802--814, 1955.

\bibitem{cox1970motion}
R.G. Cox.
\newblock The motion of long slender bodies in a viscous fluid part 1. general
  theory.
\newblock {\em J. Fluid Mech.}, 44(04):791--810, 1970.

\bibitem{pak2011}
O.~S. Pak, W.~Gao, J.~Wang, and E.~Lauga.
\newblock High-speed propulsion of flexible nanowire motors: Theory and
  experiments.
\newblock {\em Soft Matter}, 7(18):8169--8181, 2011.

\bibitem{lighthill1976flagellar}
J.~Lighthill.
\newblock Flagellar hydrodynamics.
\newblock {\em SIAM Rev.}, 18(2):161--230, 1976.

\bibitem{hancock1953self}
G.J. Hancock.
\newblock The self-propulsion of microscopic organisms through liquids.
\newblock {\em P. Roy. Soc. Lond. A Mat.}, 217:96--121, 1953.

\bibitem{keller1976slender}
J.B. Keller and S.~Rubinow.
\newblock Slender-body theory for slow viscous flow.
\newblock {\em J. Fluid Mech.}, 75(4):705--714, 1976.

\bibitem{johnson1979flagellar}
R.E. Johnson and C.J. Brokaw.
\newblock Flagellar hydrodynamics. a comparison between resistive-force theory
  and slender-body theory.
\newblock {\em Biophys. J.}, 25(1):113--127, 1979.

\bibitem{wiggins1998flexive}
C.H. Wiggins and R.E. Goldstein.
\newblock Flexive and propulsive dynamics of elastica at low reynolds number.
\newblock {\em Phys. Rev. Lett.}, 80(17):3879, 1998.

\bibitem{wiggins1998trapping}
C.H. Wiggins, D.~Riveline, A.~Ott, and R.E. Goldstein.
\newblock Trapping and wiggling: Elastohydrodynamics of driven microfilaments.
\newblock {\em Biophys. J.}, 74(2):1043--1060, 1998.

\bibitem{kantsler2012fluctuations}
V.~Kantsler and R.E. Goldstein.
\newblock Fluctuations, dynamics, and the stretch-coil transition of single
  actin filaments in extensional flows.
\newblock {\em Phys. Rev. Lett.}, 108(3):038103, 2012.

\bibitem{young2007stretch}
Y.-N. Young and M.J. Shelley.
\newblock Stretch-coil transition and transport of fibers in cellular flows.
\newblock {\em Phys. Rev. Lett.}, 99(5):058303, 2007.

\bibitem{becker2001instability}
L.E. Becker and M.J. Shelley.
\newblock Instability of elastic filaments in shear flow yields
  first-normal-stress differences.
\newblock {\em Phys. Rev. Lett.}, 87(19):198301, 2001.

\bibitem{iserles1996}
A.~Iserles.
\newblock {\em A first course in the numerical analysis of differential
  equations}.
\newblock Cambridge University Press, 1996.

\bibitem{quennouz2015transport}
N.~Quennouz, M.J. Shelley, O.~du~Roure, and A.~Lindner.
\newblock Transport and buckling dynamics of an elastic fibre in a viscous
  cellular flow.
\newblock {\em J. Fluid Mech.}, 769:387--402, 2015.

\bibitem{brennen1977}
C.~Brennen and H.~Winet.
\newblock Fluid mechanics of propulsion by cilia and flagella.
\newblock {\em Annu. Rev. of Fluid Mech.}, 9(1):339--398, 1977.

\bibitem{bigoni2012}
D.~Bigoni.
\newblock {\em Nonlinear solid mechanics: bifurcation theory and material
  instability}.
\newblock Cambridge University Press, 2012.

\bibitem{landau1986}
L.D. Landau and E.M. Lifshitz.
\newblock {\em Theory of Elasticity, vol. 7}.
\newblock Elsevier, New York, 1986.

\bibitem{guglielmini2012buckling}
L.~Guglielmini, A.~Kushwaha, E.S.G. Shaqfeh, and H.A. Stone.
\newblock Buckling transitions of an elastic filament in a viscous stagnation
  point flow.
\newblock {\em Phys. Fluids}, 24(12):123601, 2012.

\bibitem{deng2015}
M.~Deng, L.~Grinberg, B.~Caswell, and G.~E. Karniadakis.
\newblock Effects of thermal noise on the transitional dynamics of an
  inextensible elastic filament in stagnation flow.
\newblock {\em Soft matter}, 11(24):4962--4972, 2015.

\bibitem{euler1774methodus}
L.~Euler.
\newblock {\em Methodus lnveniendi Lineas Curvas Maximi Minimive Proprietate
  Gaudentes (Appendix, De curvis elasticis)}.
\newblock Lausanne and Geneva: Marcum Michaelem Bousquet, 1774.

\bibitem{timoshenko1970}
S.~Timoshenko and J.M. Gere.
\newblock {\em Theory of Elastic Stability, 2ed.}
\newblock Tata McGraw-Hill Education, 1970.

\bibitem{beck1952knicklast}
M.~Beck.
\newblock Die knicklast des einseitig eingespannten, tangential gedr{\"u}ckten
  stabes.
\newblock {\em Z. Angew. Math. Phys.}, 3(3):225--228, 1952.

\bibitem{ziegler1952}
H.~Ziegler.
\newblock Die stabilit{\"a}tskriterien der elastomechanik.
\newblock {\em Arch. Appl. Mech.}, 20(1):49--56, 1952.

\bibitem{dormand1980}
J.R. Dormand and P.J. Prince.
\newblock A family of embedded runge-kutta formulae.
\newblock {\em J. Comput. Appl. Math.}, 6(1):19--26, 1980.

\bibitem{chwang75}
A.~T. Chwang and T.~Y.~T. Wu.
\newblock Hydromechanics of low-{R}eynolds-number flow. 2. {S}ingularity method
  for {S}tokes flows.
\newblock {\em J. Fluid Mech.}, 67:787--815, 1975.

\bibitem{cortez2008}
J.~Ainley, S.~Durkin, R.~Embid, P.~Boindala, and R.~Cortez.
\newblock The method of images for regularized stokeslets.
\newblock {\em J. Comput. Phys.}, 227:4600--4616, 2008.

\bibitem{svoboda1994}
K.~Svoboda and S.~M. Block.
\newblock Force and velocity measured for single kinesin molecules.
\newblock {\em Cell}, 77(5):773--784, 1994.

\bibitem{blake74_image}
J.~R. Blake and A.~T. Chwang.
\newblock Fundamental singularities of viscous-flow. {P}art 1. {I}mage systems
  in vicinity of a stationary no-slip boundary.
\newblock {\em J. Eng. Math.}, 8:23--29, 1974.

\bibitem{gittes1993}
F.~Gittes, B.~Mickey, J.~Nettleton, and J.~Howard.
\newblock Flexural rigidity of microtubules and actin filaments measured from
  thermal fluctuations in shape.
\newblock {\em J. Cell Biol.}, 120(4):923--934, 1993.

\end{thebibliography}

\end{document}